\documentclass[floats,floatfix,nofootinbib]{revtex4}
\usepackage[dvips]{graphicx}
\usepackage{graphics}
\usepackage{amssymb}
\usepackage{bm}
\usepackage{amsmath}
\usepackage{color}
\usepackage{epstopdf}

\begin{document}

\title{Yukawa Black Holes from Interacting Vacuum }

\author{Rodrigo Maier\footnote{rodrigo.maier@uerj.br},
\vspace{0.5cm}}

\affiliation{Departamento de F\'isica Te\'orica, Instituto de F\'isica, Universidade do Estado do Rio de Janeiro,\\
Rua S\~ao Francisco Xavier 524, Maracan\~a,\\
CEP20550-900, Rio de Janeiro, Brazil\\
}

%\affiliation{
%}

\date{\today}

\begin{abstract}
In this paper we obtain an exact solution of Einstein field equations assuming an interaction 
between a vacuum component and the Maxwell field. The key feature of such interaction refers to a simple stress
exchange so that the electromagnetic field naturally incorporates the Yukawa potential.
It is shown that
the resulting spacetime thus obtained can either be a naked singularity or a black hole with 
an inner Cauchy horizon $R_-$ and an exterior event horizon $R_+$.
For this latter configuration we examine the group velocity of test photons in the region $R>R_+$.
Beyond a lower bound for the frequency
we show that superluminal velocities arise in a neighbourhood of the event horizon and that the coupling parameter of the interaction is actually connected 
to a nonvanishing rest mass 
for the photon.
\end{abstract}

\maketitle

\section{Introduction}
\label{intro}
The issue of dark energy in black hole physics has been a subject of interest over the last years\cite{mo}-\cite{javed}.
In \cite{wang}, for instance, the authors examine Kerr-Newman-AdS solutions of the Einstein-Maxwell equations considering a quintessence field around a black hole with multiple horizons.  
On the other hand, in \cite{noller} the authors study the possibility of probing the physics of a dynamical scalar field -- responsible to trigger late-time acceleration -- in black hole merger processes with gravitational wave interferometers.

From a cosmological point of view, it has been shown that an interacting dark energy component may relieve some cosmological tensions of observational data\cite{salvatelli}-\cite{kumar}. Therefore, a question which naturally arises is what would be the resulting black hole once we assume an interacting vacuum component which may play the role of dark energy. In \cite{maier} it has been shown that a Reissner-Nordstr\"om-de Sitter spacetime supports a nonsingular interior solution as long as a vacuum component properly interacts with a nonrelativistic perfect fluid. In this paper we do not take into account black hole interior solutions. In fact, we are rather interested in the case that such vacuum component interacts with an exterior matter distribution. To this end we examine one of the simplest configurations in which the exterior matter distribution is given by a Maxwell field. 

It is well known that Yukawa black holes should be one of the natural extensions of Reissner-Nordstr\"om 
spacetimes. In fact, the Yukawa gravitational potential in nonlinear electrodynamics and $f(R)$ 
theories has already been considered in literature (see \cite{halilsoy,chabab} and references therein).
Relevant astrophysical implications were shown in \cite{osorio} where the authors consider photon rings 
and magnetized discs in a neighbourhood of Yukawa black holes which emerge from
$f(R)$ gravity. 

In the framework of nuclear interaction, H. Yukawa proposed an original idea to provide
massive mesons\cite{yukawa} irrespective of electric charges. In the core of his proposal, 
such particles should be described by fields satisfying the Proca equation which admits plane wave solutions.
In this vein, it has been shown that a nonzero photon rest mass 
can be naturally incorporated into electromagnetism through a 
Proca type equation\cite{liang}.
The Yukawa potential has a broad range of applications encompassing 
several configurations in particle physics\cite{Myhrer:1987af} and condensed matter physics\cite{pel}.
In the framework of the physics of black holes the Yukawa potential is an idea extracted from the nuclear physics applied 
to modified gravity theories in order to make phenomenological predictions about gravitational potentials.
In this context, the gravitational mechanism to provide a Proca type equation for a test particle is still an open issue.

In this paper we show that Yukawa black holes with an event horizon $R_+$ may be obtained as long
as a vacuum component interacts with the Maxwell field through a judicious stress exchange.
The phenomenological motivation for such interaction relies on the attempting to furnish a nonvanishing photon rest mass\cite{greiner}. Through a Proca-like equation
we examine the group velocity of test photons and 
it turns out that -- beyond a lower bound for the frequency -- superluminal velocities arise in a neighbourhood of the event horizon.
In the asymptotic regime $R\rightarrow +\infty$, we show that the coupling parameter of the interaction is actually connected 
to a nonvanishing rest mass 
for the photon. 

We organize the paper as follows. In the next section we present an exact spherically symmetric solution of Einstein field equations assuming a proper interaction between the vacuum component and the Maxwell field so that the resulting spacetime turns 
out to be a Yukawa black hole with an exterior event horizon. In section $3$, we examine the group velocity of test photons
in this background geometry. In section $4$ we leave our final remarks.

\section{Exact Solutions}
\label{sec:1}
To start, let us consider the Einstein field equations
\begin{eqnarray}
\label{einstein}
G_{\mu\nu}-V_Ig_{\mu\nu}=-\kappa^2T_{\mu\nu},
\end{eqnarray}
where $G_{\mu\nu}$ is the Einstein tensor built with
the Christoffel symbols and $\kappa^2$ the Einstein constant. We regard $V_I$ as a vacuum component which interacts with a Maxwell field whose energy-momentum tensor reads
%
%where $T_{\mu\nu}$ the energy-momentum tensor which comes from the Maxwell field. That is,
%
\begin{eqnarray}
\label{tmunu}
T_{\mu\nu}=F_\mu^{~\alpha}F_{\nu\alpha}-\frac{1}{4}F g_{\mu\nu}.
\end{eqnarray}
Here $F$ is the Faraday scalar $F_{\mu\nu}F^{\mu\nu}$ with $F_{\mu\nu}=\nabla_\nu A_\mu-\nabla_\mu A_\nu$. 
It is well known that equations (\ref{einstein}) must be subjected to the Bianchi identities.
In the case of an interacting vacuum component, such identities are automatically satisfied as long as
\begin{eqnarray}
\label{i1}
\kappa^2 \nabla_\mu T^\mu_{~\nu}=\nabla_\nu V_I\equiv \Theta_\nu,
\end{eqnarray}
where $\Theta_\nu$ is a $4$-vector which stands for the energy-momentum exchange between the vacuum component and the Maxwell field.

For the noninteracting case it is well known that variations of the action
\begin{eqnarray}
\label{acmax}
S_M=\frac{1}{4}\int\sqrt{-g}Fd^4x,
\end{eqnarray}
with respect to $A_\alpha$, furnishes the Maxwell equations $\nabla_\mu F^{\mu\nu}=0$ which are consistent with
the Bianchi identities $\nabla_\mu T^{\mu\nu}=(\nabla_\mu F^{\mu\gamma}) F^{\nu}_{~\gamma}=0$. In the context of (\ref{i1}) however, the Bianchi identities
suggest that the Maxwell action $S_M$ must be modified in order to include the interaction between the vacuum component and the Maxwell field. Variations of such modified action with respect to the field $A_\alpha$ should be consistent with Bianchi identities
(\ref{i1}) so that the vacuum component $V_I$ should depend on $A_\alpha$ furnishing the $4$-vector $\Theta_\nu$. 
In order to probe for the microscopic origin of such energy-momentum exchange, we might consider
the action
\begin{eqnarray}
\label{ac}
S_M+\frac{1}{\kappa^2}\int\sqrt{-g}V_Id^4x.
\end{eqnarray}
Variations of this action with respect to $A_\mu$ furnishes
\begin{eqnarray}
\frac{\delta V_I}{\delta A_\mu}=\kappa^2\nabla_\nu F^{\mu\nu}.
\end{eqnarray}
From (\ref{i1}) on the other hand, we obtain
\begin{eqnarray}
\kappa^2 \nabla_\mu T^\mu_{~\nu}=\kappa^2(\nabla_\mu F^{\mu\gamma}) F_{\nu_\gamma}=\nabla_\nu V_I \rightarrow \frac{\delta V_I}{\delta A_\mu}F_{\mu\nu}=\nabla_\nu V_I
=\Theta_\nu.
\end{eqnarray}
At this stage we notice that the simplest choice for $V_I$ as a function of $A_\mu$ is $V_I\propto A_\mu A^\mu$.
If gauge invariance is abandoned, such a term in the action (\ref{ac}) has the phenomenological motivation to furnish a nonvanishing photon rest mass\cite{greiner}. 
Hence, restricting ourselves to this case we end up with
\begin{eqnarray}
\label{theta}
\Theta_\nu\propto F_{\nu\mu}A^\mu.
\end{eqnarray}
In the following we shall explore the predictions of this choice from pure
phenomenological grounds. 
%At this stage one could naively guess that the full action which furnishes
%(\ref{einstein}) and (\ref{i1}) is given by
%
%\begin{eqnarray}
%\label{fac}
%S=\frac{1}{2\kappa^2}\int\sqrt{-g}(R+2V_I)d^4x+S_M
%\end{eqnarray}
%
%with $V_I\propto A_\alpha A^\alpha$.
%
%We remark however that variations of the action (\ref{fac}) with respect to $g^{\mu\nu}$ do not furnish the field equations
%(\ref{einstein}). In fact, the piece (\ref{ac}) is rather auxiliary just to guide a choice for the 
%energy transfer between the vacuum component and the Maxwell field.

Let us then consider the case of a static and spherically symmetric geometry whose line element reads
\begin{eqnarray}
\label{mg}
ds^2=A(r)dt^2-\frac{1}{B(r)}dr^2-r^2(d\theta^2+\sin^2{\theta}d\phi^2).
\end{eqnarray}
It is well known that the Reissner-Nordstr\"om-de Sitter spacetime is a solution of (\ref{einstein}) and (\ref{i1}) for a vanishing 
$\Theta_\nu$.
In this case $A_\nu$ incorporates the Coulomb potential with overall charge $q$.
That is $A_\nu=(q/r, 0, 0, 0)$. To extend this scenario 
we leave the four potential as 
\begin{eqnarray}
\label{eqa}
A_\nu=(Q(r), 0, 0, 0).
\end{eqnarray} 
With this choice, from the field equations (\ref{einstein}) we obtain
\begin{eqnarray}
G_{tt}-G_{rr}=\frac{1}{rA}\Big(\frac{dA}{dr} B- A \frac{dB}{dr}\Big)=0 \rightarrow A(r)\equiv C B(r),
\end{eqnarray}
where $C$ is a constant of integration. Without loss of generality we shall make $C=1$.

Following the prescription (\ref{theta}) we now fix $\Theta_\nu$ as
\begin{eqnarray}
\label{eqb}
\Theta_\nu= \xi^2 \zeta^2(r)F_{\nu\alpha}A^\alpha,
\end{eqnarray}
where $\xi$ is a coupling coefficient and $\zeta^2(r)$ is just the rescaled Einstein's constant, namely, $\zeta^2(r)=\kappa^2 B(r)$. As we are working in a static spacetime and $\Theta_\nu\equiv \Theta_\nu(r)$, from now on we refer to the energy-momentum exchange $4$-vector as a stress transfer.

Substituting (\ref{tmunu}), (\ref{eqa}) and (\ref{eqb}) in (\ref{i1}), we obtain
\begin{eqnarray}
\kappa^2\frac{dQ}{dr}\Big[\xi^2 Q(r)-\frac{2}{r}\frac{dQ}{dr}-\frac{d^2Q}{dr^2}\Big]=0,\\
\kappa^2\xi^2Q(r)\frac{dQ}{dr}+\frac{dV_I}{dr}=0.
\end{eqnarray}
Assuming that $Q(r)$ is not a constant the above equations can be integrated furnishing
\begin{eqnarray}
\label{Lambda}
V_I(r)&=&\Lambda_0-\frac{\kappa^2 q_0 q_1\xi}{2r^2}-e^{-2r\xi}\Big(\frac{\kappa^2\xi^2q_0^2}{2r^2}\Big)-e^{2r\xi}\Big(\frac{\kappa^2q_1^2}{8r^2}\Big),\\
\label{pot}
Q(r)&=&e^{-r\xi}\Big(\frac{q_0}{r}\Big)+e^{r\xi}\Big(\frac{q_1}{2r\xi}\Big),
\end{eqnarray}
where $\Lambda_0$ may play the role of a cosmological constant -- in the deep IR -- and $q_0$ and $q_1$ are constants of integration. 
Finally, feeding the Einstein field equations (\ref{einstein}) with (\ref{Lambda}) and (\ref{pot}) we obtain 
\begin{eqnarray}
\nonumber
\frac{1}{r^2}-\frac{B}{r^2}-\frac{1}{r}\frac{dB}{dr}-\Lambda_0&=&\frac{\kappa^2}{8r^4\xi^2}[2q_0\xi  e^{-2r\xi}+q_1(1-2r\xi)]\\
\label{g00}
&&~~~~~~~\times[q_1 e^{2r\xi}+2q_0\xi(1+2r\xi)],\\
\nonumber
\frac{1}{r}\frac{dB}{dr}+\frac{1}{2}\frac{d^2B}{dr^2}+\Lambda_0&=&\frac{\kappa^2 e^{-2r\xi}}{8r^4 \xi^2}\Big\{4q_0\xi[q_1e^{2r\xi}+q_0\xi(1+2r\xi(1+r\xi))]\\
\label{g11}
&&~~~~~~+e^{4r\xi}q_1^2[1-2r\xi(1-r\xi)]\Big\}.
\end{eqnarray}
It is easy to see that the general solution of (\ref{g00}) and (\ref{g11}) reads
\begin{eqnarray}
\label{st0}
B(r)=1+\xi \kappa^2 q_0 q_1+\frac{\kappa^2q_0q_1}{2\xi r^2} -\frac{2GM}{r}+e^{-2r\xi}\Big(\frac{\kappa^2 q_0^2}{2 r^2}\Big)\\
\nonumber
~~~~~~~~~~~~~~~~~~~~~~~~~~~~~~~~~~~~~~~~~~~~~~~~~~+e^{2r\xi}\Big(\frac{\kappa^2 q_1^2}{8\xi^2 r^2}\Big)-\frac{\Lambda_0 r^2}{3}.
\end{eqnarray}
In order to obtain an asymptotically flat spacetime we shall restrict ourselves to $q_1=0$. For $\xi>0$ we then 
see from (\ref{pot}) that the $Q(r)$ naturally incorporates the Yukawa potential\cite{yukawa,liang} and
\begin{eqnarray}
\label{st}
B(r)=1 -\frac{2GM}{r}+e^{-2r\xi}\Big(\frac{\kappa^2 q_0^2}{2 r^2}\Big)-\frac{\Lambda_0 r^2}{3}.
\end{eqnarray}
In the limit $\xi\rightarrow 0$ we obtain the Reissner-Nordstr\"om-de Sitter spacetime -- with overall charge proportinal to $q_0$ -- as one should expect. For $\xi>0$ we see from (\ref{Lambda}) and (\ref{pot}) that (\ref{st}) inherits an exponential term which is connected to a
tail of the vacuum component. In this sense, for a proper choice of the parameters one can then obtain
a Yukawa-de Sitter black hole due to the vacuum tail described by (\ref{Lambda}). In the following we shall restrict ourselves to the case $\xi>0$.

\begin{figure}[tbp]
\begin{center}
\includegraphics[width=6.5cm,height=6cm]{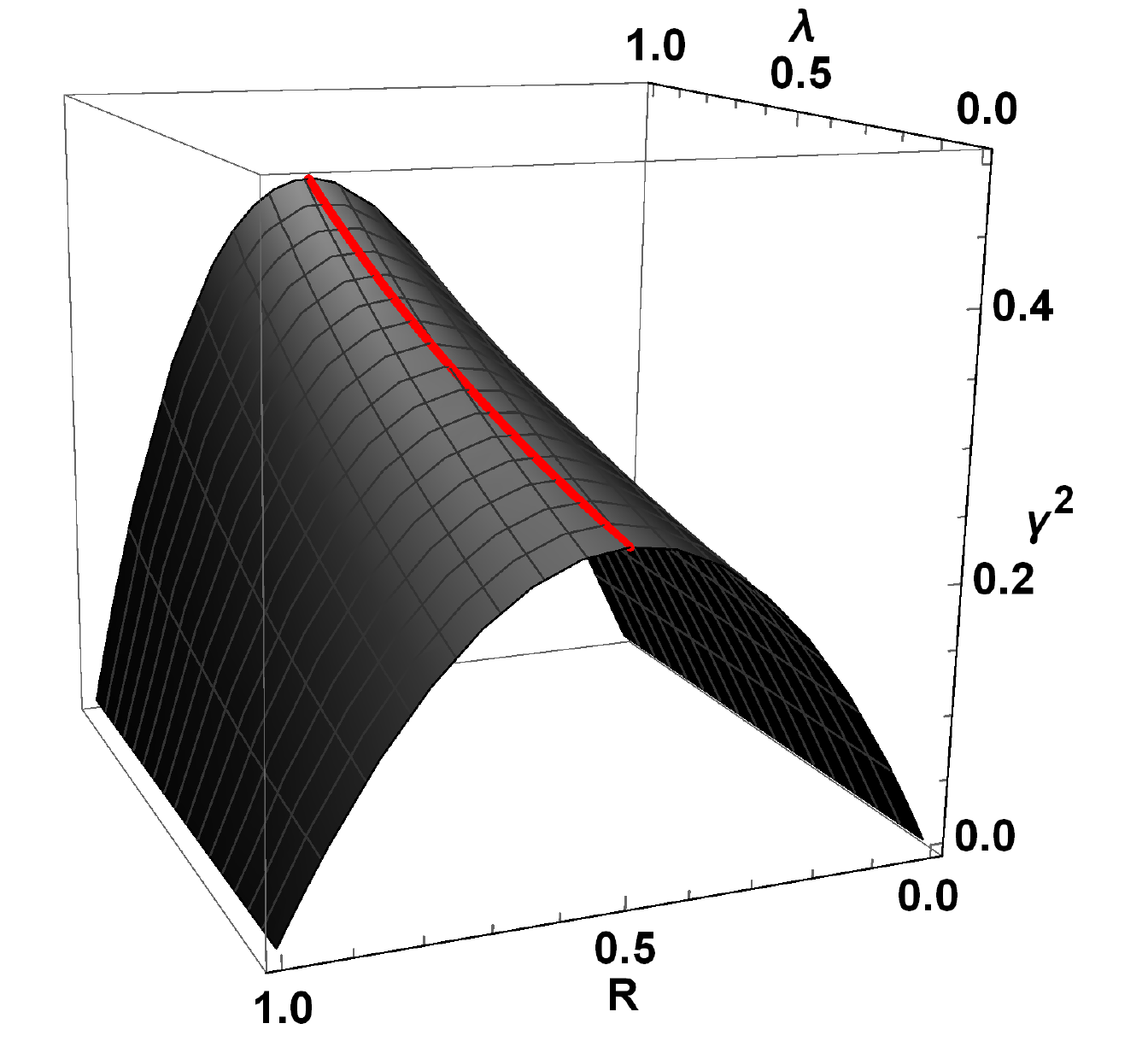}
\caption{Countour plot of the conditon $B(R)=0$ for $\lambda>0$. While the red solid curve correspond to extremal black holes,
the surface below it is connected to black holes with an exterior event horizon and an interior Cauchy horizon (see text).
}
\end{center}
\end{figure}
\begin{figure}[tbp]
\begin{center}
\includegraphics[width=5.cm,height=3cm]{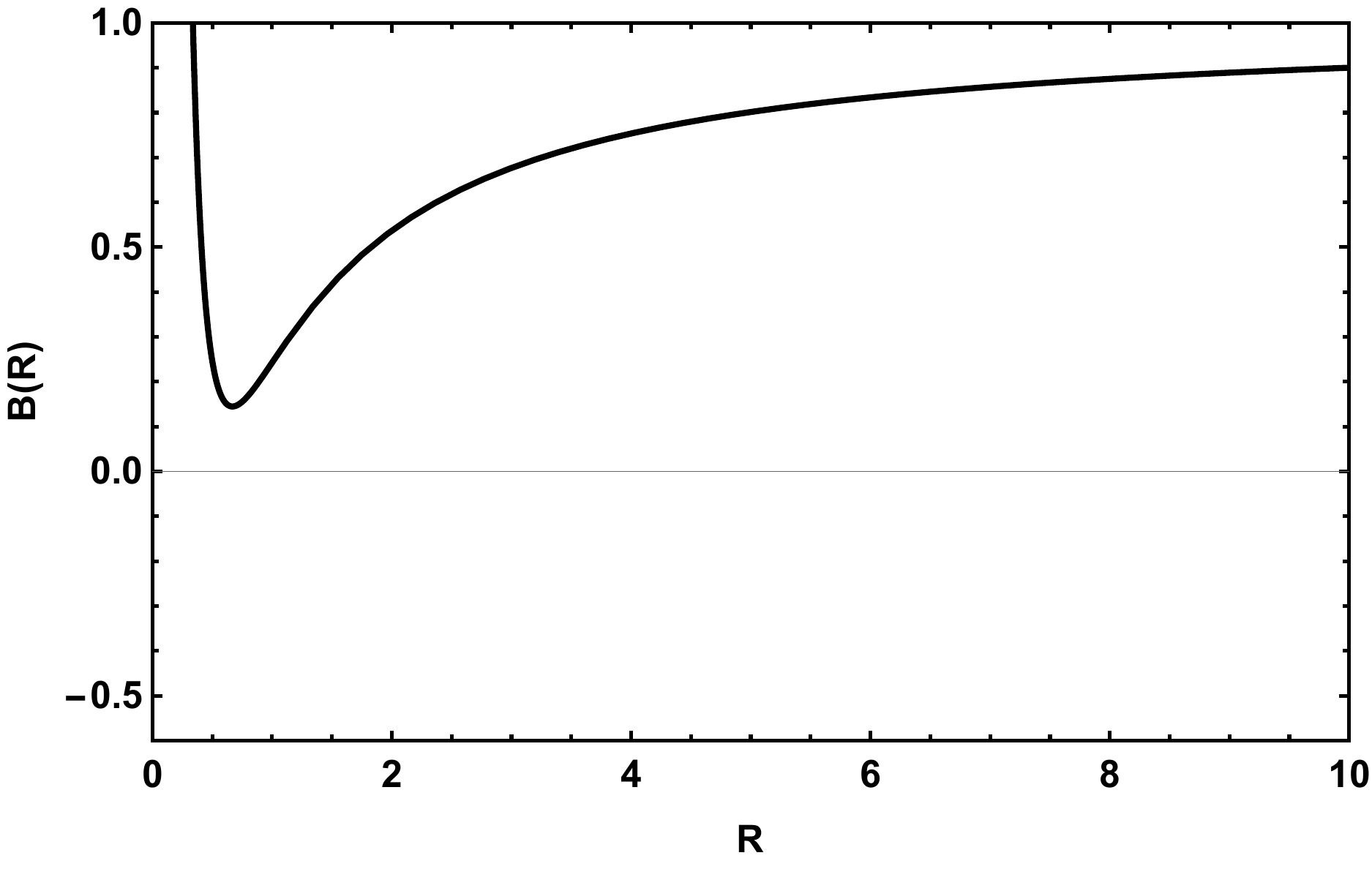}\includegraphics[width=5.cm,height=3cm]{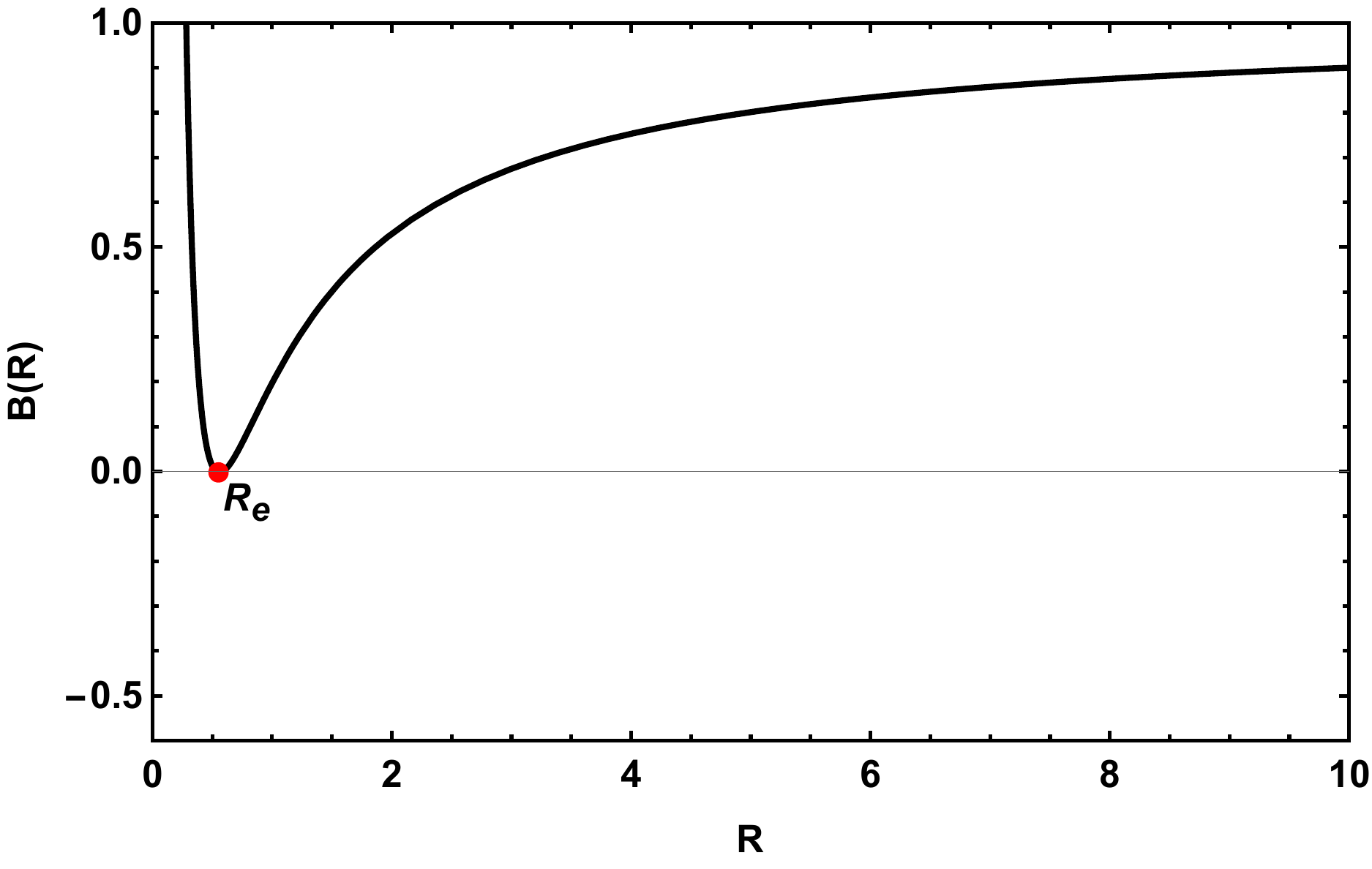}
\includegraphics[width=5.cm,height=3cm]{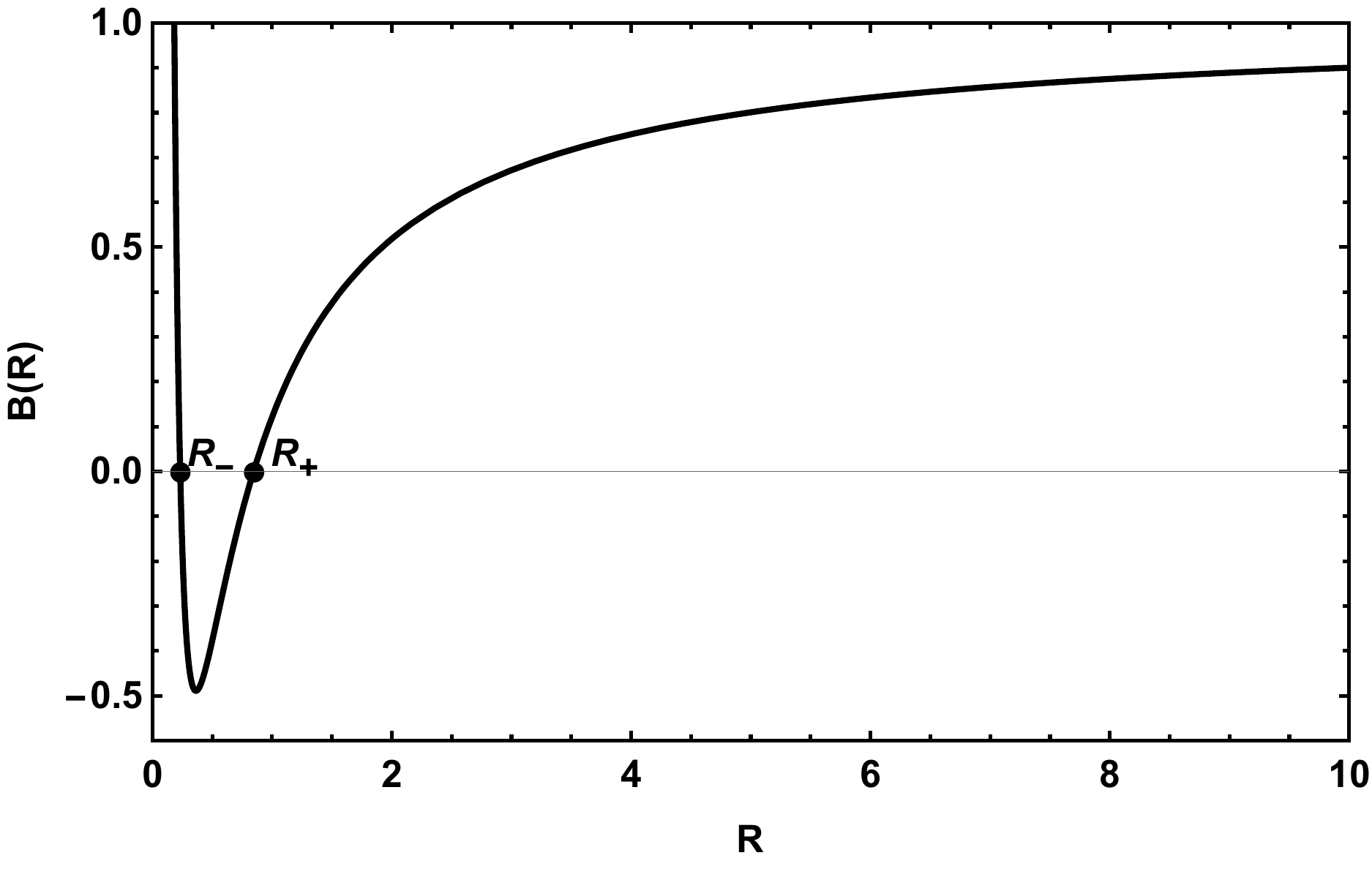}
\caption{Parametric plot of $B(R)$ for $\lambda=0.5$ and several choices of $\gamma^2$. For $\gamma^2=0.4$ (left panel)
there are no roots of $B(R)$ so that we obtain a naked singularity. For $\gamma^2\simeq 0.325$ (middle panel), $B(R)$ has only one root
connected to an extremal case. Finally, for $\gamma^2=0.2$ (right panel) we see that there are two
coordinate singularities (roots of $B(R)$) connected to an event horizon $R_+$ and a Cauchy horizon $R_-$.
}
\end{center}
\end{figure}
\begin{figure}[tbp]
\begin{center}
\includegraphics[width=6.5cm,height=8cm]{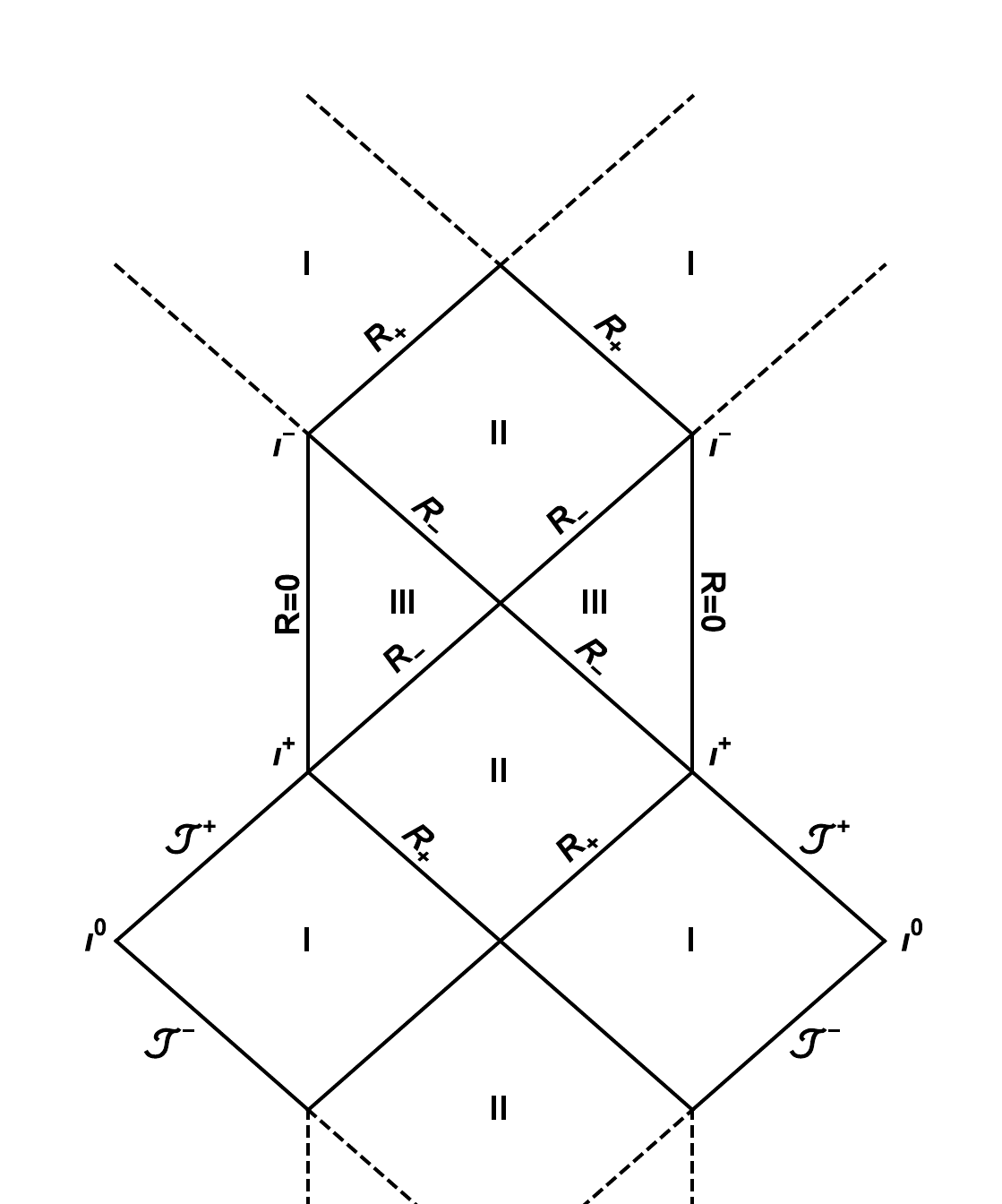}
\caption{Penrose diagram for configurations (iii). The infinite chain of asymptotically flat regions
$I$ $(\infty > R > R_+)$ are connected to regions $III$ $(R_- > R > 0)$ by regions $II$ $(R_+ > R > R_-)$.}
\end{center}
\end{figure}

To examine the presence of horizons we will set $\Lambda_0\rightarrow 0$ to simplify our analysis.
Regarding $\Lambda_0$ as a cosmological constant, this turns out to be a reasonable assumption once an asymptotic dark energy component should dominate only at a low energy scale
in order to drive late-time acceleration. In that case, the solution (\ref{st}) reads
\begin{eqnarray}
\label{src}
B(R)=1-\frac{1}{R}+\frac{\gamma^2}{R^2}e^{-\lambda R}
\end{eqnarray}
in rescaled units
\begin{eqnarray}
\label{res}
R=\frac{r}{2GM},~~\lambda=4GM\xi,~~\gamma=\frac{\kappa q_0}{2\sqrt{2}GM}.
\end{eqnarray}
For $\lambda> 0$ it can be shown that (\ref{src}) furnishes three different configurations: (i) naked
singularities, (ii) extremal black holes with an horizon $R_e$ or (iii) a black hole with a Cauchy horizon $R_-$ and an exterior event horizon
$R_+$.  
In fact, from (\ref{src}) we see that 
\begin{eqnarray}
\lim_{R\rightarrow 0^+}B(R)=+\infty~~{\rm and}~~\lim_{R\rightarrow +\infty}B(R)=1.
\end{eqnarray}
Therefore, the condition for horizon formation may be satisfied as long as $B(R)$ has at least one global minimum.
In order to seek for extremal configurations, let us consider the equations
\begin{eqnarray}
\label{re1}
B(R_e)=0,\\
\label{re2}
\frac{dB}{dR}\Big|_{R_e}=0.
\end{eqnarray}
From (\ref{re1}) it follows that
\begin{eqnarray}
\label{g2}
\exp(-R_e \lambda)=\frac{R_e}{\gamma^2}(1-R_e).
\end{eqnarray}
Substituting this result in (\ref{re2}) we end up with
\begin{eqnarray}
\label{re}
R_e=\frac{1}{2}-\frac{1}{\lambda}+\frac{\sqrt{\lambda^2+4}}{2\lambda},
\end{eqnarray}
for $R_e>0$. We finally note that $\gamma^2$ as a function of $\lambda$ can be easily obtained from (\ref{g2}) and (\ref{re})
so that extremal configurations satisfy
the following parametrization in the $3$-dimensional space $(\lambda, R, \gamma^2)$:
\begin{eqnarray}
\label{par}
\Phi(\lambda)=\Big(\lambda, \frac{1}{2}-\frac{1}{\lambda}+\Delta, \frac{2}{\lambda^2}(\lambda\Delta-1)e^{-1+\lambda(1/2+
 \Delta)}\Big),
\end{eqnarray}
where
$\Delta\equiv {\sqrt{\lambda^2+4}}/{2\lambda}$.

In Fig. 1 we illustrate the above parametrization corresponding to extremal configurations by the red solid curve.
For a fixed pair $(\lambda, \gamma^2)$ above this curve there are no real solutions for the equation $B(R)=0$. Such a region is connected
naked singularities.
Finally, below the red solid curve we illustrate a surface connected to the condition $B(R)=0$. This surface refers to configurations (iii). 
In fact, for a fixed pair $(\lambda, \gamma^2)$ below $\Phi(\lambda)$, the condition $B(R)=0$ furnishes two possible values for coordinates singularities,
one corresponding to an external event horizon $R_+$ and the other corresponding to a Cauchy horizon $R_-$ -- analogous to that
of a Reissner-Nordstr\"om black hole. In Fig. 2 we illustrate one example of each of these configurations fixing $\lambda=0.5$
and different choices of $\gamma^2$. As $B(R)\rightarrow 1$ as $R\rightarrow \infty$, we note that $B(R)>0$ for $R>R_+$.
The maximal analytical extension for configurations (iii) is illustrated by the Penrose diagram
in Fig. 3.

The presence of the Cauchy horizon $R_-$ in the
maximal analytical extension of the geometry (\ref{src}) may pose the question of a
possible instability of the spacetime. In order to make a first inceptive analysis
about this issue in the present context,
we consider possible instabilities as measured by geodetic observers as
in \cite{gursel,Matzner:1979zz}. In this vein, let us consider a massless test scalar field
$\Psi$ in the background described by (\ref{st}) with $\Lambda_0=0$. 
By assumption, such scalar field satisfies the wave equation 
\begin{eqnarray}
\label{psi1}
\Box \Psi=0.
\end{eqnarray}
Taking the Fourier transform and considering background symmetries, one may expand $\Psi$ in spherical harmonics
as
\begin{eqnarray}
\label{psi2}
\Psi(t, r, \theta, \phi)=\sum_{l,m}\int_{-\infty}^{+\infty}\frac{\psi_{l m \omega}{(r)}}{r}e^{-i\omega t}Y_{lm} (\theta, \phi)d\omega.
\end{eqnarray}
Substituting (\ref{psi2}) in (\ref{psi1}) we then obtain
\begin{eqnarray}
\frac{d^2\psi_{l m \omega}}{dr^2_\ast}+[\omega^2-U(r)]\psi_{l m \omega}=0,
\end{eqnarray}
where 
\begin{eqnarray}
r_\ast=\int\frac{1}{B(r)}dr.
\end{eqnarray}
In the above, 
\begin{eqnarray}
\label{pots}
U(r)=\frac{B}{r}\Big[\frac{l(l+1)}{r^2}+\frac{dB}{dr}\Big]
\end{eqnarray}
stands for the scattering potential -- a basic feature which references \cite{gursel,Matzner:1979zz} rely on.

Assuming a negligible coupling parameter, it is easy to see that
\begin{eqnarray}
B(r)\simeq 1- \frac{2GM+\kappa^2 q_0^2\xi}{r}+\frac{\kappa^2q_0^2}{2r^2}.
\end{eqnarray}
up to first order in $\xi$. In this case,
\begin{eqnarray}
r_\pm\simeq GM+\frac{\kappa^2q_0^2\xi}{2}\pm\sqrt{(2GM+\kappa^2q_0^2\xi)^2-2\kappa^2q_0^2},
\end{eqnarray}
so that we may rewrite
\begin{eqnarray}
\label{bap}
B(r)\simeq\frac{(r-r_+)(r-r_-)}{r^2}.
\end{eqnarray}
Substituting (\ref{bap}) in (\ref{pots}) we obtain the very same structure of the scattering potential used in Refs.
\cite{gursel,Matzner:1979zz} for the case of a Reissner-Norstr\"om spacetime. In this case, possible instabilities 
would arise on $R_-$ since
the energy density of the scalar field develops
singularities in a neighborhood of the Cauchy horizon (as shown in \cite{gursel,Matzner:1979zz}).
It is true that this analysis holds as long 
as $\xi$ is a negligible small parameter. In the following we will show that $\xi$ is actually connected to the
photon rest mass -- a feature which justifies our argument in the present context.
%Overall, 
%one should expect a possible instability on $R_-$ since
%the energy density of the scalar field develops
%singularities in a neighborhood of the Cauchy horizon (as shown in \cite{gursel,Matzner:1979zz}).
%once in this approximation the scattering potential is the same as in \cite{gursel,Matzner:1979zz},.

\section{The Photon Rest Mass}
\label{sec:2}

To proceed we now consider the equations of motion of test photons in the background described by
the solution (\ref{src}). To this end, let us assume a Maxwell test field whose lagrangian reads
\begin{eqnarray}
\label{at}
S_p=\frac{1}{4}\int \sqrt{-g}{\cal F}d^4x,
\end{eqnarray}
where ${\cal F}$ is the Faraday scalar ${\cal F}_{\mu\nu}{\cal F}^{\mu\nu}$
with ${\cal F}_{\mu\nu}=\nabla_\nu {\cal A}_\mu-\nabla_\mu {\cal A}_\nu$. Assuming that
such Maxwell test field interacts with the vaccum component in the same manner as the electromagnetic field
of our background solution, we end up with the following extended Proca equations\cite{proca}
\begin{eqnarray} 
\label{ep1}
\nabla_\mu {\cal F}^{\mu\alpha}-\xi^2 \zeta^2(R){\cal A}^\alpha=0.
\end{eqnarray}
Equations (\ref{ep1}) can be rewritten in terms of the vector potential ${\cal A}_\alpha$ so that we obtain
\begin{eqnarray}
\label{ep2}
\Big[\Box + \xi^2\zeta^2(R)\Big]{\cal A}_\alpha=R_{\alpha\sigma}{\cal A}^\sigma+\nabla_\alpha(\nabla_\sigma {\cal A}^\sigma).
\end{eqnarray} 
%
%where we have imposed the Lorentz condition $\nabla_\mu {\cal A}^\mu\equiv 0$.
The quest for a complete solution of (\ref{ep2}) in the curved background (\ref{src}) is a rather involved task 
which we intend to address in
a further analysis. In the present context instead we are interested in the behaviour of test photons
in a background described by (\ref{src}) in the domain $R>R_+$ of configurations (iii).
 
%in the exterior region of the event horizon $R_+$. 
%In order to simplify our analysis we will also fix $\Lambda_0\equiv 0$ again. This
%turns out to be a reasonable assumption once a cosmological constant poses a huge problem to quantum field
%theory on how to accommodate its observed value with vacuum energy calculations\cite{wein}.
%Moreover, from a theoretical point of view there is a plethora of 
%alternatives to explain a late-time accelerated regime (see \cite{amen} and Refs. therein) which are connected to asymptotic %flat solutions. 
%

Let us then assume a photon described by a test field given by the approximation
\begin{eqnarray}
{\cal A}_\alpha=C_\alpha e^{i k_\mu x^\mu},
\end{eqnarray}
where $C_\alpha$ are constants of amplitude and $k_\mu=(\omega, -{\bf k})$.
Imposing the Lorentz condition $\nabla_\mu {\cal A}^\mu\equiv 0$ and fixing
$C_\alpha=(C_t, C_r, 0, 0)$, it can be seen that equation (\ref{ep2}) reduces to
\begin{eqnarray}
\omega^2=B^2(R)\Big({\bf k}^2+\xi^2 -\frac{1}{R B}\frac{dB}{dR}-\frac{1}{2B}\frac{d^2B}{dR^2}\Big),
\end{eqnarray}
in units $c=\kappa^2=1$.
Therefore, the group velocity of free waves would then take the form
\begin{eqnarray}
\label{vp}
v_g=\frac{d\omega}{dk}= B(R)\sqrt{1-\frac{\Delta(R)}{\omega^2}},
\end{eqnarray}
where
\begin{eqnarray}
\Delta(R)=B(R)\Big[\xi^2 B(R)-\frac{1}{R}\frac{dB}{dR}-\frac{1}{2}\frac{d^2B}{dR^2}\Big]. 
\end{eqnarray}
\begin{figure}[tbp]
\begin{center}
\includegraphics[width=7.cm,height=5cm]{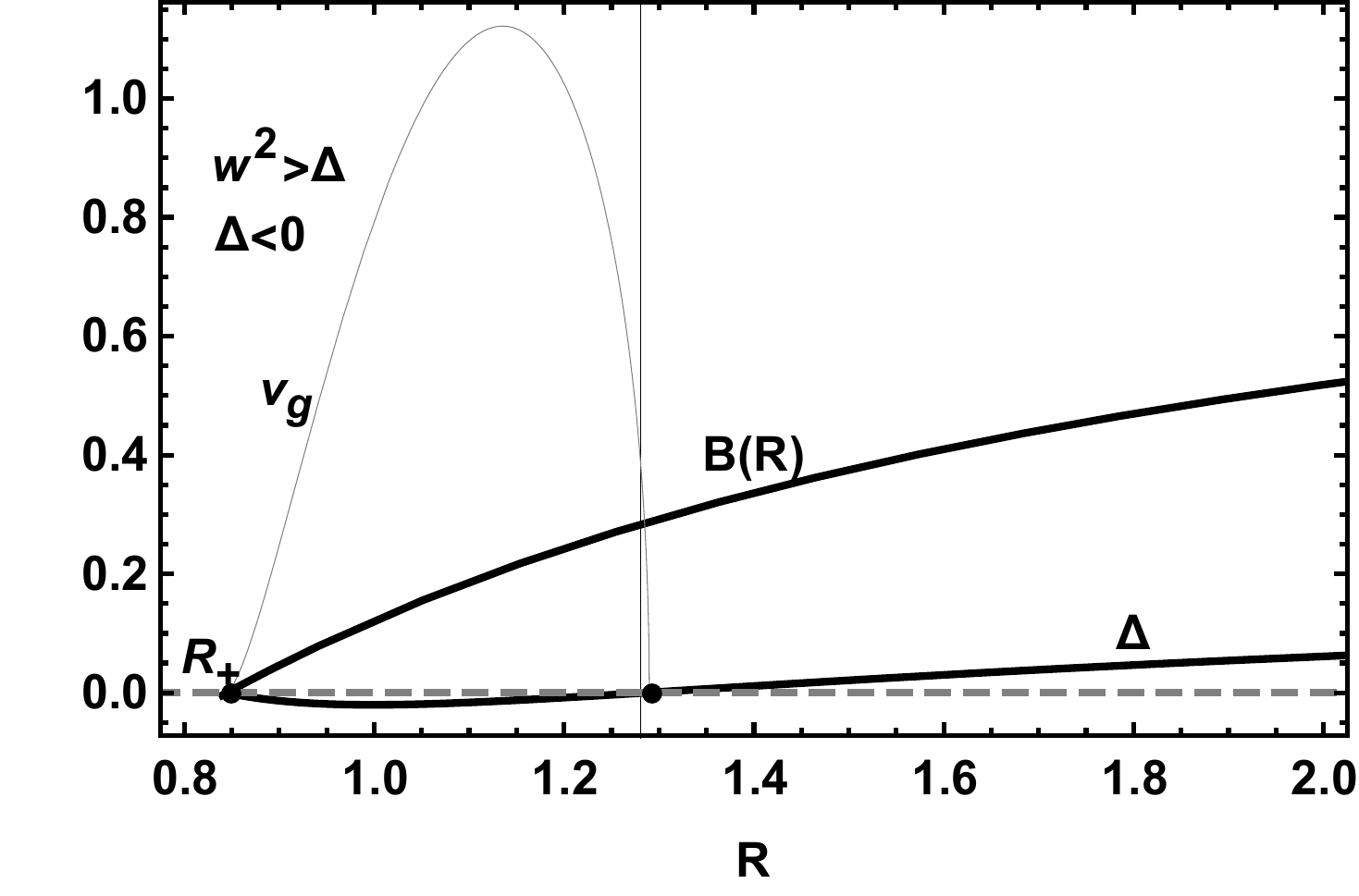}\includegraphics[width=7.cm,height=5cm]{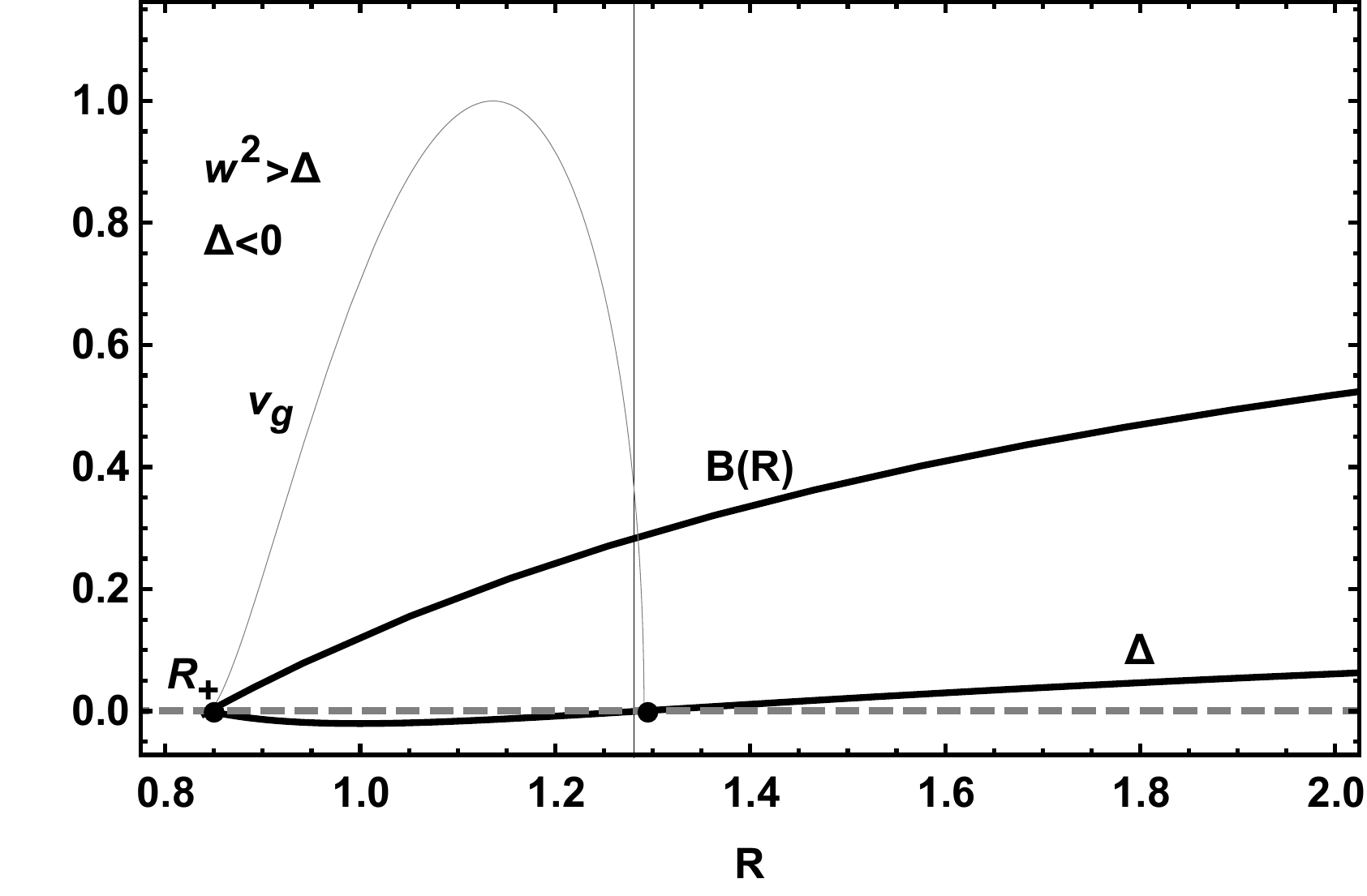}\\\includegraphics[width=7.cm,height=5cm]{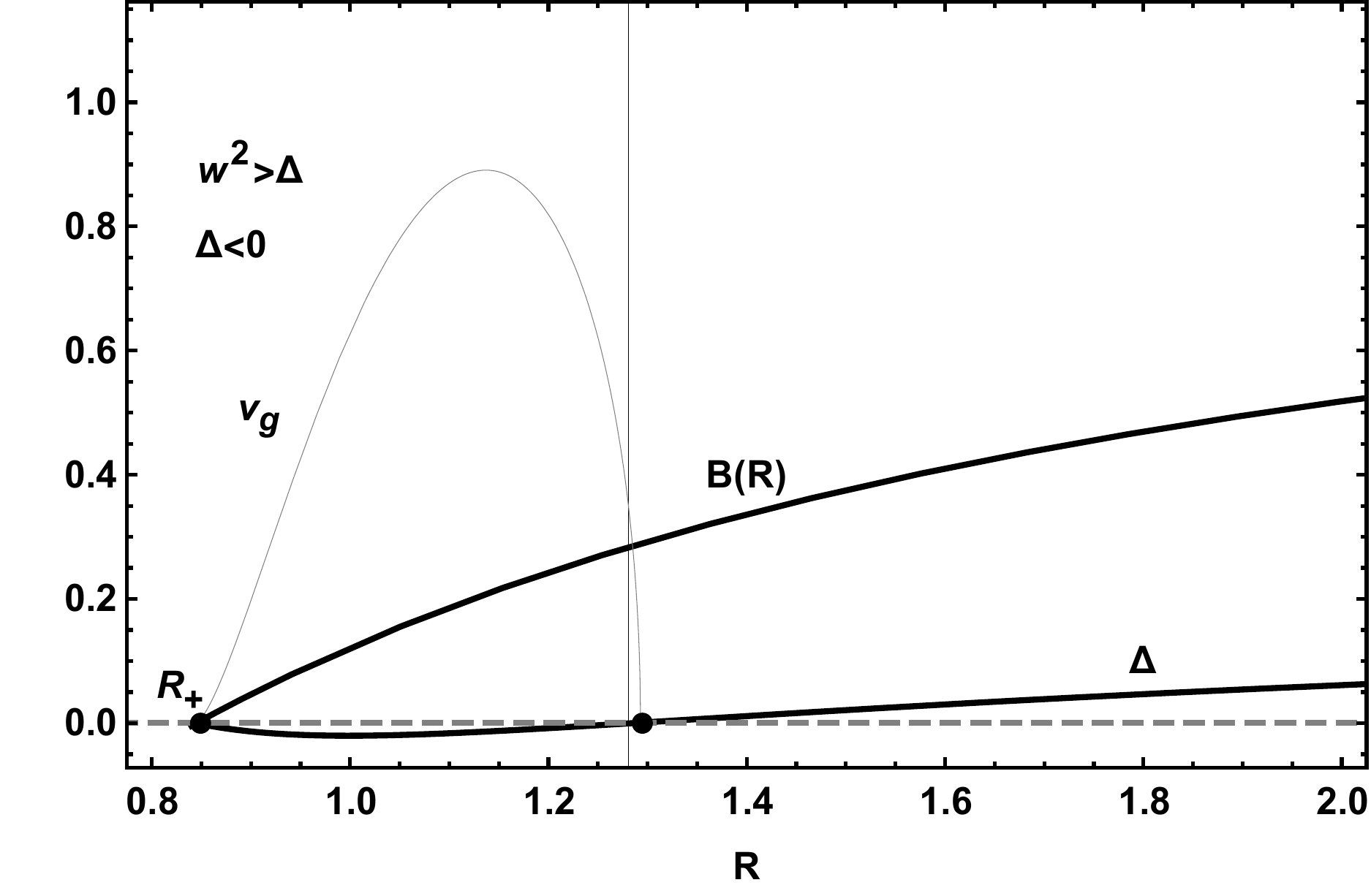}
\caption{The group velocity for $\gamma^2=0.2$, $\lambda=0.5$.
In each panel we consider different values for $\omega$, namely $\omega=0.0220$ (top left), 
$\omega=0.0248$ (top right) and $\omega=0.0280$ (bottom). 
Gray dashed horizontal lines stand for each numerical value of $\omega$.
Vertical lines located at $R\simeq 1.281$ stand for $\Delta=0$.
The region on the left of the vertical lines
are connected to the domain $\omega^2>\Delta$ and $\Delta<0$. Here $v_g>B(R)$ and free waves may propagate with superluminal velocity
in a finite domain of $R$ (see top left panel). From the above we see that there is a lower bound of $\omega$ (namely, $\omega=0.0248$, top right panel) beyond which
the group velocity is superluminal.
In a neighbourhood of the right side of the vertical lines (where $\omega^2>\Delta>0$) we see that $v_g$ decreases 
rapidly (with $v_g<B(R)$) until it vanishes at the bold black dots illustrated. Beyond these black dots, $\omega^2<\Delta$ and the
group velocity (and $k$)
becomes imaginary. 
}
\end{center}
\end{figure}

From (\ref{vp}) it is straightforward to see that $v_g\rightarrow 0$ as $R \rightarrow R_+$ so that free waves freeze at the event horizon.
Moreover, for $\lambda>0$ ($\xi>0$), from (\ref{src}) we note that
\begin{eqnarray}
-\frac{1}{R}\frac{dB}{dR}-\frac{1}{2}\frac{d^2B}{dR^2}=-\frac{\gamma^2 e^{-\lambda R}(2+\lambda R(2+\lambda R))}{2R^4}<0,
\end{eqnarray}
so that $\Delta(r)$ can be either positive or negative according to a given domain of $R$ and the coupling parameter $\xi$ -- 
remember that $B(R)>0$ for $R>R_+$ as discussed in the previous section. Therefore we then note
that there are four domains of interest: (i) for $\omega^2>\Delta$ and $\Delta<0$  
we see that $v_g>B(R)$ so that there is a lower bound for $\omega$ beyond which
free waves propagate with superluminal velocities; (ii)  when $\omega^2>\Delta>0$ free waves propagate
with a velocity of energy flow smaller than $B(R)$ and approaches to it as the frequency tends to infinity; 
(iii) for $\omega^2=\Delta$ 
free waves do not propagate at all, that is $v_g=0$; (iv)
when $\omega^2<\Delta$ the group velocity (and $k$)
becomes imaginary and the amplitude of free massive waves would be attenuated exponentially.
In Fig. 4 we illustrate these domains of interest considering the particular case shown in Fig. 2 (right panel).

It is easy to see that superluminal configurations are absent in
the asymptotic regime $B(R)\rightarrow 1$. In fact, in this case
\begin{eqnarray}
\label{vpa}
v_g\rightarrow \sqrt{1-\frac{\xi^2}{\omega^2}},
\end{eqnarray}
and 
%the above configurations reduce to: (i) 
when $\omega>\xi$ massive waves propagate
with a velocity of energy flow smaller than $1$ (or $c$) and approaches to it as the frequency tends to infinity.
In this sense we see from (\ref{vpa}) that the coupling coefficient $\xi$ arises as an asymptotic photon rest mass due to the interaction between the Maxwell test field with the vacuum component.

%For the last but not least we now address the gauge invariance of our model. To start, it is easy to see that 
%the electromagnetic tensor in (\ref{tmunu}) is invariant through the transformation
%
%\begin{eqnarray}
%\label{g1}
%A_\alpha\rightarrow A_\alpha + \Pi_{,\alpha}, 
%\end{eqnarray}
%
%where $\Pi$ is an arbitrary scalar function. Nevertheless, gauge invariance is automatically broken once we assume a stress %transfer of the type (\ref{eqb}). In order to circumvent this issue one may rewrite the vector potential in terms of a 
%Stueckelberg field\cite{stu} $S$, namely $\mathbb{A}_\nu=A_\nu - S_{,\nu}$, so that 
%$F_{\nu\alpha}\rightarrow  \nabla_\alpha\mathbb{A}_\nu-\nabla_\nu\mathbb{A}_\alpha$ and $\Theta_\nu\rightarrow \xi^2 %\zeta^2(r)F_{\nu\alpha}\mathbb{A}^\alpha$.
%
%
%Therefore, applying the extended gauge transformation\cite{pauli}
%
%\begin{eqnarray}
%\label{g2}
%A_\alpha\rightarrow A_\alpha + \Pi_{,\alpha}~~{\rm and}~~ S \rightarrow S+\Pi,
%\end{eqnarray}
%
%we guarantee that gauge invariance is manifest in the field equations of our model. In this context
%the Yukawa interaction which appears in (\ref{pot}) and ({\ref{src}}) can be analogously realised. Furthermore, performing the 5same procedure for the test Maxwell field, a nonvanishing rest mass
%for the photon still asymptotically arises as a consequence of the interaction between the vacuum component
%and the Maxwell field. 

\section{Final Remarks}
\label{sec:2}

In the context of General Relativity we have shown how Yukawa black holes 
may be obtained once one assumes a judicious choice for the interaction 
between a vacuum component and the Maxwell field. The key feature of such interaction refers to a simple 
term $F_{\mu\nu} A^\mu$ -- inspired by
an attempted to furnish a nonvanishing photon rest mass\cite{greiner} -- so that the electromagnetic field naturally incorporates the Yukawa potential.
The resulting spacetime turns out to be either a naked singularity or a black hole with an 
interior Cauchy horizon $R_-$ and an exterior event horizon $R_+$.
For test photons in such background we present a mechanism in which a Proca-like equation arises.
We examine the group velocities of such photons in the particular $R>R_+$.
We show that beyond a lower bound for the frequency, superluminal velocities arise in a neighbourhood of the event horizon.
In the asymptotic regime $R\rightarrow +\infty$
a nonvanishing rest mass 
for the photon arises as a consequence 
of the interaction. 

We remark that the solutions found in this paper are rather phenomenological since the stress
exchange lack of a complete and proper microphysical motivation. The complete quest of microphysical grounds to sustain such stress exchange will be a subject of a further publication.
%We address the gauge invariance of our model assuming that
%the electromagnetic vector potential supports
%a Stueckelberg field.

As a future perspective we should point out that the complete examination of (\ref{ep2}) in curved spacetime will be a subject for a further publication.
Another issue which deserves a careful examination is what are the backreaction consequences of such test massive photons 
in the background described by (\ref{src}).  

For the last but not least, we should mention that the exact analysis about the Cauchy horizon $R_-$ 
in the maximal analytical extension of the geometry (\ref{src}) may pose the question about
instabilities of the spacetime as measured by geodetic observers\cite{gursel}. 
In fact, the Cauchy horizon is a global boundary in which the field equations lose
their power to describe the evolution of prior initial conditions. It has been shown that
for a free falling observer crossing the Cauchy horizon, an arbitrary large blue-shift of
any incoming radiation would be seen so that the flux of energy of test fields would
diverge once crossing it (see Ref. \cite{chandra} and references therein). In this sense, the Cauchy
horizon is a null surface of infinite blue-shift.
It is well known that this may constitute a problem so that we intend to approach
the complete analysis about this issue in a future publication.

\end{document}